\documentclass[twocolumn]{revtex4}
\usepackage{mathbbold}
\usepackage{amsfonts}
\usepackage{amsmath}
\usepackage{amssymb}
\usepackage{charter}
\usepackage{graphicx}

\begin{document}

\title{Thermal state truncation by using quantum scissors device}
\author{Hong-xia Zhao$^{1}$, Xue--xiang Xu$^{2,\dag }$, Hong-chun Yuan$^{3}$}
\affiliation{$^{1}$Information Engineering college, Jiangxi University of technology,
Nanchang 330098, China \\
$^{2}$Center for Quantum Science and Technology, Jiangxi Normal University,
Nanchang 330022, China\\
$^{3}$College of Electrical and Optoelectronic Engineering, Changzhou
Institute of Technology, Changzhou 213002, China\\
$^{\dag }$Corresponding author: xxxjxnu@gmail.com }

\begin{abstract}
A non-Gaussian state being a mixture of the vacuum and single-photon states
can be generated by truncating a thermal state in a quantum scissors device
of Pegg et al. [Phys. Rev. Lett. 81 (1998) 1604]. In contrast to the thermal
state, the generated state shows nonclassical property including the
negativity of Wigner function. Besides, signal amplification and
signal-to-noise ratio enhancement can be achieved.


\end{abstract}

\maketitle

\section{Introduction}

The generation of quantum states is a prerequisite for universal quantum
information processing (QIP) \cite{1}. Quantum states are usually classified
into discrete-variable (DV) and continuous-variable (CV) descriptions \cite%
{2}. In the CV quantum regime, there are two classes of quantum states that
play an important role in QIP: Gaussian and non-Gaussian states, referring
to their character of wave function or Wigner function \cite{3,4}. In
general, Gaussian states are relatively easy to generate and manipulate
using current standard optical technology\cite{5}.

However, in the recent several decades, some probabilistic schemes are
proposed to generate and manipulate non-Gaussian states \cite{6,6a,6b}. Many
schemes work in postselection \cite{7}, that is, the generated state is
accepted conditionally on a measurement outcome. The typical examples
include photon addition and subtraction \cite{8}, and noise addition \cite{9}%
. Among them, an interesting scheme was based on the quantum-scissors
devices. In 1998, Pegg, Phillips and Barnett proposed this quantum state
truncation scheme, which change an optical state $\gamma _{0}\left\vert
0\right\rangle +\gamma _{1}\left\vert 1\right\rangle +\gamma _{2}\left\vert
2\right\rangle +\cdots $ into qubit optical state $\gamma _{0}\left\vert
0\right\rangle +\gamma _{1}\left\vert 1\right\rangle $. The device is then
called a quantum scissors device (QSD), while the effect is referred to as
optical state truncation via projection synthesis. This quantum mechanical
phenomenon was actually a nonlocal effect relying on entanglement because no
light from the input mode can reach the output mode \cite{10}. After its
proposal, an experiment of quantum scissors was realized by Babichev, Ries
and Lvovsky \cite{11} by applying the experimentally feasible proposal of
Ref. \cite{11a1,11a2,11a3}. The QSD was also applied and generalized to
generate not only qubits but also qutrits \cite{11b} and qudits \cite%
{11c1,11c2} of any dimension. Similar quantum state can be also generated
via a four-wave mixing process in a cavity \cite{11d}.

Following these works on QSD, Ferreyrol et al. implemented a
nondeterministic optical noiseless amplifier for a coherent state \cite{12}.
Moreover, heralded noiseless linear amplifications were designed and
realized \cite{13,14,15}. Recently, an experimental demonstration of a
practical nondeterministic quantum optical amplification scheme was
presented to achieve amplification of known sets of coherent states with
high fidelity \cite{16}. By the way, many systems transmitting signals using
quantum states could benefit from amplification. In fact, any attempt to
amplify signal must introduce noise inevitably. In other words, perfect
deterministic amplification of an unknown quantum signal is impossible. In
addition, Miranowicz et. al. studied the phase-space interference of quantum
states optically truncated by QSD \cite{16a}.

Inspired by the above works, we generate a non-Gaussian mixed state by using
a Gaussian thermal state as the input state of the quantum scissors in this
paper. This process transform an input thermal state into an incoherent
mixture of only zero-photon and single-photon components. The success
probability of such event is studied. Some properties of the generated
state, such as signal amplification, signal-to-noise ratio and the
negativity of the Wigner function, are investigated in detail. The paper is
organized as follows. In section II, we outline the framework of QSD and
introduce the scheme of thermal state truncation. Quantum state is derived
explicitly and the probability is discussed. Subsequently, some statistical
properties, such as average photon number, intensity gain, signal-to-noise
ratio, are investigated in section III. In addition, we study the Wigner
function and the parity for the output state in section IV. Conclusions are
summarized in the final section.

\section{Thermal state truncation scheme}

In this section, we outline the basic framework of quantum scissors
device and introduce our scheme of thermal state truncation.

\subsection{Framework of \textbf{quantum scissors device}}

QSD mainly includes two beam splitters (BSs) and three channels, as shown in
Fig.1. Three channels are described by the optical modes $a$, $b$, and $c$
in terms of their respective creation (annihilation) operators $a^{\dag }$($%
a $), $b^{\dag }$($b$) and $c^{\dag }$($c$). Since every channel have an
input port and an output port, the QSD have six ports. The interaction
including several key stages as follows. Firstly, the channel $a$ and the
channel $c$\ are correlated through an asymmetrical beam splitter (A-BS),
whose operation can be described by the unitary operator $B_{1}=e^{\theta
\left( a^{\dag }c-ac^{\dag }\right) }$ with the transmissivity $T=\cos
^{2}\theta $. After that, the channel $b$ and the channel $c$\ are then
correlated through another symmetrical beam splitter (S-BS, also 50:50 BS),
whose operation can be described by the unitary operator $B_{2}=e^{\frac{\pi
}{4}\left( b^{\dag }c-bc^{\dag }\right) }$. Moreover, among these six ports,
four ports are fixed with special processes as follows: (1) Injecting the
auxiliary single-photon $\left\vert 1\right\rangle $ in the input port of
channel $a$; (2) Injecting the auxiliary zero-photon $\left\vert
0\right\rangle $ in the input port of channel $c$; (3) Detecting the
single-photon $\left\vert 1\right\rangle $ in the output port of channel $b$%
; and (4) Detecting the zero-photon $\left\vert 0\right\rangle $ in the
output port of channel $c$.

QSD leaves only one input port (i.e., the input port in channel $b$) and one
output port (i.e., the output port in channel $a$). Injecting an appropriate
input state in the input port, one can generate a new quantum state in the
output port. Many previous theoretical and experimental schemes have used
the pure states as the input states to generated quantum states. Here, our
proposed scheme use a mixed state as the input state to generate quantum
state.

\subsection{Thermal state truncation}

Using a mixed state (i.e., thermal state) as the input state, we shall
generate another mixed state in our present protocol. The input thermal
state is given by%
\begin{equation}
\rho _{th}=\sum_{n=0}^{\infty }\frac{\bar{n}^{n}}{\left( \bar{n}+1\right)
^{n+1}}\left\vert n\right\rangle \left\langle n\right\vert ,  \label{1}
\end{equation}%
where $\bar{n}$\ is the average number of the thermal photons \cite{17}.
Therefore, the output generated state can be expressed as%
\begin{eqnarray}
\rho _{out} &=&\frac{1}{p_{d}}\left\langle 0_{c}\right\vert \left\langle
1_{b}\right\vert B_{2}\{\rho _{th}\otimes   \notag \\
&&[B_{1}(\left\vert 1_{a}\right\rangle \left\langle 1_{a}\right\vert \otimes
\left\vert 0_{c}\right\rangle \left\langle 0_{c}\right\vert )B_{1}^{\dag
}]\}B_{2}^{\dag }\left\vert 1_{b}\right\rangle \left\vert 0_{c}\right\rangle
\label{2}
\end{eqnarray}%
where $p_{d}$ is the success probability.
\begin{figure}[tbp]
\label{Fig1} \centering\includegraphics[width=0.9\columnwidth]{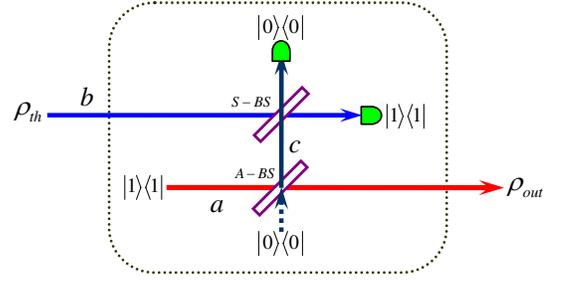}
\caption{(Colour online) Conceptual scheme of "quantum scissors device"
(QSD) for thermal state truncation. The auxiliary single-photon $\left\vert
1\right\rangle \left\langle 1\right\vert $ in channel $a$ and the auxiliary
single-photon $\left\vert 0\right\rangle \left\langle 0\right\vert $ in
channel $c$ generates an entangled state between the modes $a$ and $c$\
after passing through an asymmetrical beam splitter (A-BS) with the
transmissivity $T$. The input mode $b$ (accompanied by the input thermal
state $\protect\rho _{th}$) is then combined with $c$ in a (50:50)
symmetrical beam splitter (S-BS). A successful heralded truncation
(accompanied by the output generated state $\protect\rho _{out}$) in the
output $a$ mode is flagged by a single-photon event in the output $b$ mode
detection and no photons on the output $c$ mode detection.}
\end{figure}

The explicit density operator in Eq.(\ref{2}) can further be expressed as%
\begin{equation}
\rho _{out}=p_{0}\left\vert 0\right\rangle \left\langle 0\right\vert
+p_{1}\left\vert 1\right\rangle \left\langle 1\right\vert ,  \label{3}
\end{equation}%
where $p_{0}=\left( 1-T\right) \left( \bar{n}+1\right) /\left( \bar{n}%
+1-T\right) $ and $p_{1}=\bar{n}T/\left( \bar{n}+1-T\right) $ are,
respectively, the zero-photon distribution probability and the one-photon
distribution probability. Obviously, the output state is an incoherent
mixture of a vacuum state $\left\vert 0\right\rangle \left\langle
0\right\vert $ and a one-photon state $\left\vert 1\right\rangle
\left\langle 1\right\vert $ with certain ratio coefficients $p_{0}$, $p_{1}$%
. If $T=0$, then $\rho _{out}\rightarrow \left\vert 0\right\rangle
\left\langle 0\right\vert $; while for $T=1$, then $\rho _{out}\rightarrow
\left\vert 1\right\rangle \left\langle 1\right\vert $.

From another point of view, the output generated state in Eq.(\ref{3})
remains only the first two terms of the input thermal state in Eq.(\ref{1}),
which can also be considered as an truncation from the input thermal state.
However, the corresponding coefficients of these terms are changed.
Moreover, the output generated state carry the information of the input
thermal state because it also depend on the thermal parameter $\bar{n}$.
Since no light from the input port reaches the output port, this process
also mark the nonlocal quantum effect of the operation for the quantum
scissors.

From present protocol, we easily obtain $p_{d}$ as follows
\begin{equation}
p_{d}=\allowbreak \frac{\bar{n}+1-T}{2\left( \bar{n}+1\right) ^{2}}.
\label{4}
\end{equation}%
For a given $\bar{n}$, it can be shown that $p_{d}$ is a linear decreasing
function of $T$.
\begin{figure}[tbp]
\label{Fig2} \centering\includegraphics[width=1.0\columnwidth]{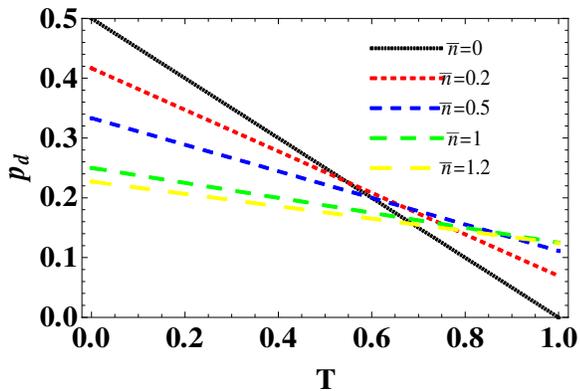}
\caption{(Colour online) Probability of successfully generating the output
state as a function of the beam-splitter transmissivity according to the
model presented in the text. The average photon number of the input thermal
state $\bar{n}$ has been fixed to 0, 0.2, 0.5, 1, 1.2.}
\end{figure}

In Fig.2, we plot $p_{d}$ as a function of $T$\ for different $\bar{n}$. For
instance, when $\bar{n}=1$, we have $p_{d}|_{\bar{n}=1}=0.25-0.125T$ (see
the green line in Fig.2); when $\bar{n}=0$, we have $p_{d}|_{\bar{n}%
=0}=\allowbreak 0.5-0.5T$ (see the black line in Fig.2). The results on the
success probability provide a theoretical reference for experimental
realization.

\section{Statistical properties of the generated state}

By adjusting the interaction parameters, i.e., the thermal parameter $\bar{n}
$ of the input state and the transmission parameter $T$ of the A-BS, one can
obtain different output states with different figures of merits. Some
statistical properties, such as average photon number, intensity gain and
signal-to-noise ratio, are studied in this section. As the reference, we
will compare the properties of the output state with those of the input
state.

\subsection{Average photon number and intensity gain}

Using the definition of the average photon number, we have $\left\langle
\hat{n}\right\rangle _{\rho _{th}}=\bar{n}$ for the input thermal state and%
\begin{equation}
\left\langle \hat{n}\right\rangle _{\rho _{out}}=\frac{\bar{n}T}{\bar{n}+1-T}%
.  \label{5}
\end{equation}%
for the output generated state. Here $\hat{n}$\ is the operator of the
photon number \cite{18}.

In Fig.3, we plot $\left\langle \hat{n}\right\rangle _{\rho _{out}}$ as a
function of $T$\ for different $\bar{n}$. Two extreme cases, such as, e.g.
(1) $\left\langle \hat{n}\right\rangle _{\rho _{out}}\equiv 0$ if $\bar{n}=0$
or $T=0$, and (2) $\left\langle \hat{n}\right\rangle _{\rho _{out}}\equiv 1$
if $T=1$ for any $\bar{n}\neq 0$, are always hold. No matter how large the
input thermal parameter $\left\langle \hat{n}\right\rangle _{\rho _{th}}$
is, there always exists $\left\langle \hat{n}\right\rangle _{\rho _{out}}\in
\lbrack 0,1]$. Moreover, $\left\langle \hat{n}\right\rangle _{\rho _{out}}$
is an increasing function of $T$ for a given nonzero $\bar{n}$.
\begin{figure}[tbp]
\label{Fig3} \centering\includegraphics[width=1.0\columnwidth]{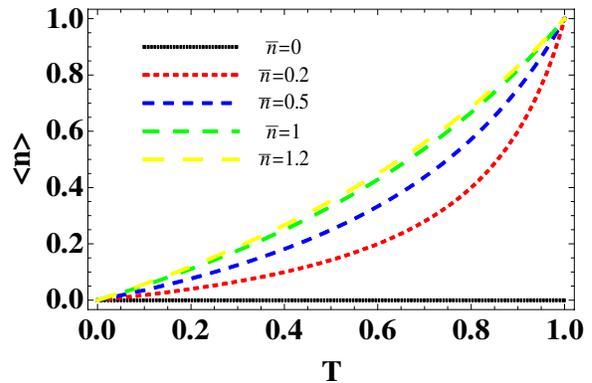}
\caption{(Colour online) Average photon number of the output state as a
function of the beam-splitter transmissivity. The average photon number of
the input thermal state $\bar{n}$ has been fixed to 0, 0.2, 0.5, 1, 1.2.}
\end{figure}

In order to describe signal amplification, we define the intensity gain as $%
g=\left\langle \hat{n}\right\rangle _{\rho _{out}}/\left\langle \hat{n}%
\right\rangle _{\rho _{th}}$, which is related with the intensity $%
\left\langle \hat{n}\right\rangle _{\rho _{out}}$ of the output field with
that ($\left\langle \hat{n}\right\rangle _{\rho _{th}}$) of the input field.
Therefore we have%
\begin{equation}
g=\allowbreak \frac{T}{\bar{n}+1-T}.  \label{6}
\end{equation}%
If $g>1$, then there exist signal amplification.
\begin{figure}[tbp]
\label{Fig4} \centering\includegraphics[width=1.0\columnwidth]{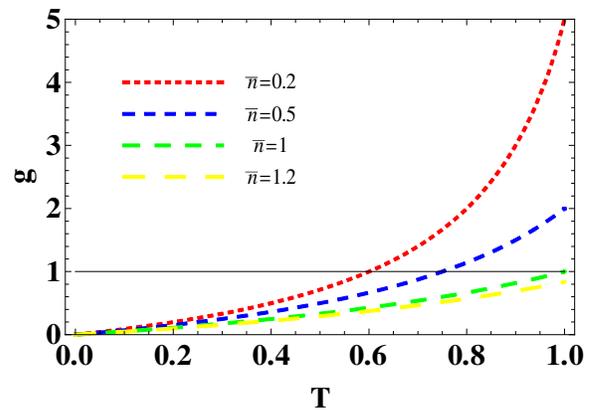}
\caption{(Colour online) Intensity gain of the output state as a function of
the beam-splitter transmissivity. The average photon number of the input
thermal state $\bar{n}$ has been fixed to 0.2, 0.5, 1, 1.2. Here the
amplification (i.e., $g>1$) occur only for $\bar{n}<1$ and $T>\left( \bar{n}%
+1\right) /2$. There exist no amplification for high-intensity ($\bar{n}>1$)
thermal state.}
\end{figure}

Fig.4 shows the intensity gain $g$ as a function of $T$ for different $\bar{n%
}$. If $\bar{n}\geq 1$, $g$ is impossible to exceed 1, which means no
amplification. In other words, the amplification happens only for the cases $%
\bar{n}<1$ with $T\in (\left( \bar{n}+1\right) /2,1]$.

\subsection{Signal to noise ratio}

Signal-to-noise ratio (abbreviated SNR or S/N) is a measure used in science
and engineering that compares the level of a desired signal to the level of
background noise \cite{19}. Here we are interesting to the effect that the
process has on the noise of these states. Typically this is shown by
calculating the variance of the photon number and forming the SNR, defined
by $SNR=\left\langle \hat{n}\right\rangle /\sqrt{\left\langle \hat{n}%
^{2}\right\rangle -\left\langle \hat{n}\right\rangle ^{2}}$. From the
definition, we have $\left\langle \hat{n}\right\rangle |_{\rho _{th}}=\bar{n}
$, $\left\langle \hat{n}^{2}\right\rangle |_{\rho _{th}}=\bar{n}+2\bar{n}^{2}
$, and then $SNR|_{\rho _{th}}=\bar{n}/\sqrt{\bar{n}+\bar{n}^{2}}$ for the
input thermal state $\rho _{th}$. While for our generated state $\rho _{out}$%
, we find $\left\langle \hat{n}\right\rangle |_{\rho _{out}}=\left\langle
\hat{n}^{2}\right\rangle |_{\rho _{out}}=p_{1}$ and%
\begin{equation}
SNR|_{\rho _{out}}=\sqrt{\allowbreak \frac{\bar{n}T}{\left( 1-T\right)
\left( \bar{n}+1\right) }}.  \label{7}
\end{equation}%
\begin{figure}[tbp]
\label{Fig5} \centering\includegraphics[width=1.0\columnwidth]{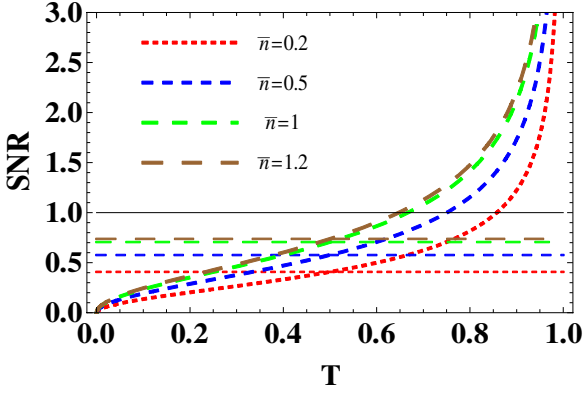}
\caption{(Colour online) Signal to noise ratio of the output states (curve
line) as a function of the beam-splitter transmissivity, compared to their
corresponding thermal states (straight line). Here $\bar{n}$ has been fixed
to 0.2, 0.5, 1, 1.2. A clear enhancement for a given $\bar{n}$ can be seen
for values of $T>0.5$. the SNR higher than 1:1 (black line) can be found in
lager $T>\left( \bar{n}+1\right) /\left( 2\bar{n}+1\right) $.}
\end{figure}

As is shown in Fig.5, we see the SNR for the output states, as compared to
their corresponding thermal states, of the same fixed average photon number.
It is found that a clear enhancement (corresponding to its input thermal
state) can be seen for $T>0.5$. Moreover, although the SNR of the input
thermal state is always smaller than 1:1, the SNR higher than 1:1 for the
output generated state can be found in larger $T$ $\left( >\left( \bar{n}%
+1\right) /\left( 2\bar{n}+1\right) \right) $.

\section{Wigner function and Parity of the generated state}

The negative Wigner function is a witness of the nonclassicality of a
quantum state \cite{20,21,22}. For a single-mode density operator $\rho $,
the Wigner function in the coherent state representation $\left\vert
z\right\rangle $ can be expressed as $W(\beta )=\frac{2e^{2\left\vert \beta
\right\vert ^{2}}}{\pi }\int \frac{d^{2}z}{\pi }\left\langle -z\right\vert
\rho \left\vert z\right\rangle e^{-2\left( z\beta ^{\ast }-z^{\ast }\beta
\right) }$, where $\beta =\left( q+ip\right) /\sqrt{2}$. Therefore we easily
obtain $W_{\rho _{th}}(\beta )=2/\left( \pi \left( 2\bar{n}+1\right) \right)
e^{-2\left\vert \beta \right\vert ^{2}/\left( 2\bar{n}+1\right) }$ for the
input thermal state and%
\begin{equation}
W_{\rho _{out}}(\beta )=p_{0}W_{\left\vert 0\right\rangle \left\langle
0\right\vert }(\beta )+p_{1}W_{\left\vert 1\right\rangle \left\langle
1\right\vert }(\beta )  \label{8}
\end{equation}%
for the output generated state with $W_{\left\vert 0\right\rangle
\left\langle 0\right\vert }(\beta )=\frac{2}{\pi }e^{-2\left\vert \beta
\right\vert ^{2}}$ \ and $W_{\left\vert 1\right\rangle \left\langle
1\right\vert }(\beta )=\frac{2}{\pi }(4\left\vert \beta \right\vert
^{2}-1)e^{-2\left\vert \beta \right\vert ^{2}}$.

As we all know, the thermal state is a Gaussian state, whose Wigner function
have no negative region. However, our output generated states have lost the
Gaussian characters because of the non-Gaussian forms of their Wigner
functions.\ In addition, the Wigner function will exhibit negative in some
region satisfying the following condition $\left\vert \beta \right\vert
^{2}<[2T\bar{n}-\left( \bar{n}+1-T\right) ]/\left( 4\bar{n}T\right) $. In
Fig.6, we plot the Wigner functions of the output generated states for two
different cases, where the negative region is found for case with large $T$.
\begin{figure}[tbp]
\label{Fig6} \centering\includegraphics[width=1.0\columnwidth]{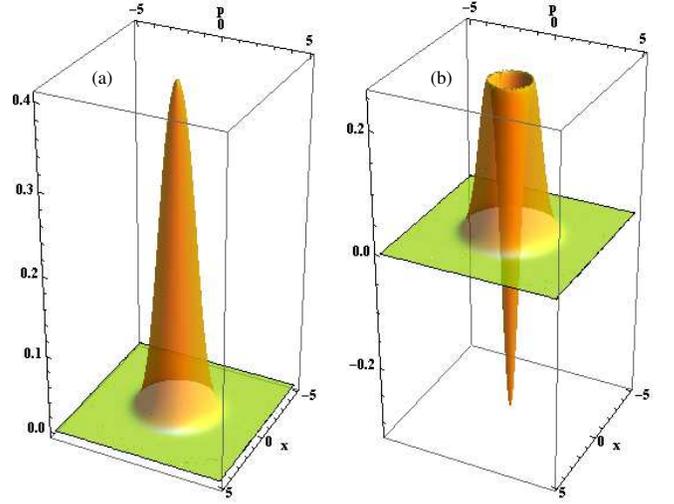}
\caption{(Colour online) Wigner function of the output state with (a) $\bar{n%
}=0.5$, $T=0.4$\ and (b) $\bar{n}=0.5$, $T=0.9$. Case (b) has the negaive
region.}
\end{figure}

Since the Wigner function of the output state is symmetrical in $x$ and $p$
space, one can determine the fact\ whether the Wigner function have negative
region by seeing $W_{\rho _{out}}(\beta =0)$. As Gerry pointed out that the
Wigner function at the origin is the expectation value of the parity
operator $\Pi =\left( -1\right) ^{\hat{n}}$, that is $\left\langle \Pi
\right\rangle =\frac{\pi }{2}W(0)$ \cite{23}. Thus, we have $\left\langle
\Pi \right\rangle _{\rho _{th}}=1/\left( 2\bar{n}+1\right) $ for the input
thermal state and%
\begin{equation}
\left\langle \Pi \right\rangle _{\rho _{out}}=\frac{\bar{n}+1-T-2T\bar{n}}{%
\allowbreak \bar{n}+1-T},  \label{9}
\end{equation}%
for the output generated state. Fig.7 show $\left\langle \Pi \right\rangle
_{\rho _{out}}$ as a function of $T$ for different $\allowbreak \bar{n}$.
\begin{figure}[tbp]
\label{Fig7} \centering\includegraphics[width=1.0\columnwidth]{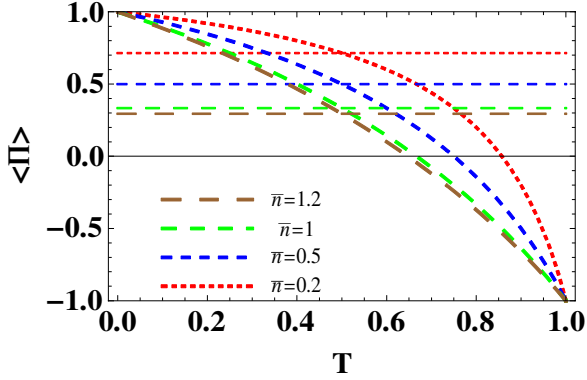}
\caption{(Colour online) Parity of the output state \textbf{as a function of
the beam-splitter transmissivity}, where $\bar{n}$ has been fixed to 0.2,
0.5, 1, 1.2. The negative region of the Wigner function can be seen for
values of $T>\left( \bar{n}+1\right) /\left( 1+2\bar{n}\right) $.}
\end{figure}

Photon number states are assigned a parity of $+1$ if their photon number is
even and a parity of $-1$ if odd \cite{24}. According to Eq.(\ref{3}), we
verify $\left\langle \Pi \right\rangle _{\rho _{out}}=p_{0}-p_{1}$. If the
condition $T>\left( \bar{n}+1\right) /\left( 1+2\bar{n}\right) $ is hold,
then there exist $\left\langle \Pi \right\rangle _{\rho _{out}}<0$, which
means that the Wigner function must exhibit negative region in the phase
space.

\section{Conclusion}

In summary, we have applied the QSD of Pegg, Philips and Barnett to truncate
a thermal field to a completely mixed qubit state, i.e., a mixture of the
vacuum and single-photon state. The explicit expression was derived in
Schrodinger picture and the success probability of such event was discussed.
The output generated state depend on two interaction parameters, i.e., the
input thermal parameter and the transmissivity of the A-BS. It is shown that
the success probability is a linear decreasing function of the
transmissivity for any given input parameter. Some nonclassical properties
of the qubit state were analyzed including intensity amplification,
singal-to-noise ratio, and the non-positive Wigner function. It was shown
that the average photon number of the output state can be adjusted between 0
and 1. The intensity amplification will happen only for small-intensity
thermal field ($\allowbreak \bar{n}<1$) and large-transmissivity ($T>\left(
\bar{n}+1\right) /2$). The SNR of the output state can be enhanced by the
operation for a given input thermal state at larger values of $T$ ($>0.5$).
The SNR higher than unity can be found in the range of $T$ $>\left( \bar{n}%
+1\right) /\left( 2\bar{n}+1\right) $. In addition, the negativity of the
Wigner function appears only for proper $T>\left( \bar{n}+1\right) /\left(
1+2\bar{n}\right) $.

\begin{acknowledgments}
We would like to thank Li-yun Hu and Bi-xuan Fan for their great helpful
discussions. This work was supported by the Research Foundation of the
Education Department of Jiangxi Province of China (Nos. GJJ151150 and
GJJ150338) and the Natural Science Foundation of Jiangxi Province of China
(20151BAB202013) as well as the National Natural Science Foundation of
China(Grants No. 11264018 and No. 11447002).
\end{acknowledgments}

\textbf{Appendix A: Derivation of the density operator in Eq.(\ref{3})}

In this appendix, we provide a detailed process of deriving the explicit
expression of the output generated state in Schrodinger picture.
Substituting $\left\vert 1_{a}\right\rangle =\frac{d}{ds_{1}}e^{s_{1}a^{\dag
}}\left\vert 0_{a}\right\rangle |_{s_{1}=0}$, $\left\langle 1_{a}\right\vert
=\frac{d}{dh_{1}}\left\langle 0_{a}\right\vert \exp e^{h_{1}a}|_{h_{1}=0}$, $%
\left\vert 1_{b}\right\rangle =\frac{d}{ds_{2}}e^{s_{2}b^{\dag }}\left\vert
0_{b}\right\rangle |_{s_{2}=0}$, $\left\langle 1_{b}\right\vert =\frac{d}{%
dh_{2}}\left\langle 0_{b}\right\vert e^{h_{2}b}|_{h_{2}=0}$, as well as%
\begin{equation*}
\rho _{th}=\frac{1}{\bar{n}}\int \frac{d^{2}\alpha }{\pi }e^{-\left( \frac{1%
}{\bar{n}}+1\right) \left\vert \alpha \right\vert ^{2}}e^{\alpha b^{\dag
}}\left\vert 0_{b}\right\rangle \left\langle 0_{b}\right\vert e^{\alpha
^{\ast }b},
\end{equation*}%
into Eq.(\ref{2}), we have%
\begin{eqnarray*}
\rho _{out} &=&\frac{d^{4}}{\bar{n}p_{d}ds_{1}dh_{1}dh_{2}ds_{2}} \\
&&\int \frac{d^{2}\alpha }{\pi }e^{-\left( \frac{1}{\bar{n}}+1\right)
\left\vert \alpha \right\vert ^{2}}\left\langle 0_{c}\right\vert
\left\langle 0_{b}\right\vert e^{h_{2}b} \\
&&e^{s_{1}\allowbreak ta^{\dag }+\frac{\alpha -s_{1}r}{\sqrt{2}}b^{\dag }-%
\frac{\alpha +s_{1}r}{\sqrt{2}}c^{\dag }}\left\vert 0_{a}\right\rangle
\left\vert 0_{b}\right\rangle \left\vert 0_{c}\right\rangle \\
&&\left\langle 0_{c}\right\vert \left\langle 0_{b}\right\vert \left\langle
0_{a}\right\vert e^{\allowbreak h_{1}ta+\allowbreak \frac{\alpha ^{\ast
}-h_{1}r}{\sqrt{2}}b-\frac{\alpha ^{\ast }+h_{1}r}{\sqrt{2}}c} \\
&&e^{s_{2}b^{\dag }}\left\vert 0_{b}\right\rangle \left\vert
0_{c}\right\rangle |_{s_{1}=s_{2}=h_{1}=h_{2}=0}
\end{eqnarray*}%
where we have used the following transformations%
\begin{eqnarray*}
B_{1}aB_{1}^{\dag } &=&at-cr,\text{ \ }B_{1}cB_{1}^{\dag }=ar+ct, \\
B_{2}bB_{2}^{\dag } &=&\frac{b-c}{\sqrt{2}},\text{ \ }B_{2}cB_{2}^{\dag }=%
\frac{b+c}{\sqrt{2}}.
\end{eqnarray*}%
and $B_{1}\left\vert 0_{a}\right\rangle \left\vert 0_{c}\right\rangle
=\left\vert 0_{a}\right\rangle \left\vert 0_{c}\right\rangle $, $%
B_{2}\left\vert 0_{b}\right\rangle \left\vert 0_{c}\right\rangle =\left\vert
0_{b}\right\rangle \left\vert 0_{c}\right\rangle $, as well as their
conjugations. In addition, $t=\cos \theta $ and $r=\sin \theta $\ are\ the
transmission coefficient and the reflection coefficient\ of the A-BS,
respectively. After detailed calculation, we obtain
\begin{eqnarray*}
\rho _{out} &=&\frac{d^{4}}{\left( \bar{n}+1\right)
p_{d}ds_{1}dh_{1}dh_{2}ds_{2}} \\
&&e^{\frac{\bar{n}}{2\left( \bar{n}+1\right) }s_{2}h_{2}-\frac{r}{\sqrt{2}}%
\left( h_{1}s_{2}+rh_{2}s_{1}\right) } \\
&&e^{s_{1}ta^{\dag }}\left\vert 0_{a}\right\rangle \left\langle
0_{a}\right\vert e^{\allowbreak h_{1}ta}|_{s_{1}=s_{2}=h_{1}=h_{2}=0}
\end{eqnarray*}%
Using $\left\vert 0_{a}\right\rangle \left\langle 0_{a}\right\vert =$ $%
:e^{-a^{\dag }a}:$ and making the derivative in the normal ordering form
(denoted by $:\cdots :$), we have%
\begin{equation*}
\rho _{out}=:\left( p_{0}+p_{1}a^{\dag }a\right) \exp \left( -a^{\dag
}a\right) :
\end{equation*}%
Thus the density operator in Eq.(\ref{3}) is obtained.

\end{document}